\documentclass[a4paper,11pt]{article}
\usepackage{jcappub} % for details on the use of the package, please
\usepackage[T1]{fontenc} % if needed
\usepackage{subfig}
\usepackage{breqn}
\usepackage{siunitx}
\def\be{\begin{equation}}
\def\ee{\end{equation}}
\def\barr{\begin{array}}
\def\earr{\end{array}}
\def\bea{\begin{eqnarray}}
\def\eea{\end{eqnarray}}
\def\bfig{\begin{figure}}
\def\efig{\end{figure}}

\title{Test of Kerr-Sen metric with black hole observations}

\author[a,b]{Ashish Narang,}
\author[a]{Subhendra Mohanty,}
\author[a]{Abhass Kumar}

\affiliation[a]{Physical Research Laboratory, Ahmedabad, 380009, India}
\affiliation[b]{Indian Institute of Technology, Gandhinagar, 382355, India}

\emailAdd{ashish@prl.res.in}
\emailAdd{mohanty@prl.res.in}
\emailAdd{abhass@prl.res.in}

\abstract{The Kerr-Sen black hole is a rotating charged black hole solution arising from heterotic string theory. In 4-dimensions effective theory the bosonic fields are: a $U(1)$ gauge boson, a Kalb-Ramond 3-form which is equivalent to a pseudoscalar axion in 4-dimensions, the dilaton and the graviton. The coupling constants in the theory are $\alpha^{\prime}$ (inverse string tension) and $\kappa$ (inverse reduced Planck mass in 4-dimensions) and the charge of the $U(1)$ field and the axion-photon coupling are related to these two. Sen found a black hole solution (the Kerr-Sen black hole) with these fields as the external hair of the black hole. In this paper we investigate the possibility of determining the Sen solution from observations.

The observations which can test the Kerr-Sen black hole are: (a) determination of the shape of the photon shadow, and (b) the rotation of polarization of photon due to axion hair. The deviation from circularity gives the $U(1)$ charge of the black hole and identification of this charge in terms of the photon coupling leads to a prediction of frequency independent ``Faraday rotation'' in terms of black hole parameters already determined from the shadow. Similar measurements of Kerr-Newman black hole with axion hair have no correlation between the shape of the image and the amount of ``Faraday rotation''. This correlation can be a distinctive test of the Sen metric.

In the recent observation from EHT of M87* shadow, the deviation from circularity has an upper bound of 10\%. If this observation is refined to 1\% accuracy then a definitive prediction of the charge of the Kerr-Sen black hole and the ``Faraday rotation'' can be made. Interestingly observations of ``Faraday rotation'' have shown that the effect is independent of frequency pointing to an axionic hair interpretation for the effect.   \\
}


\begin{document} 
\maketitle
\flushbottom

\section{Introduction}
%Ever since the first solution to general relativity obtained by Schwarzschild leading to a black hole \cite{Schwarzschild:1916uq, Finkelstein:1958zz}.

Black holes arise as solutions of various theories of gravity, notably general relativity and string theory. Their existence as physical objects has now been proven by observations of gravitational waves generated due to the merging of black holes by the LIGO experiment \cite{Abbott:2016blz, LIGOScientific:2018mvr} and also through the extraordinary observation of the M87* black hole (BH) shadow by the Event Horizon Telescope (EHT) \cite{Akiyama:2019cqa,Akiyama:2019eap,Akiyama:2019fyp,Akiyama:2019bqs,Akiyama:2019brx,Akiyama:2019sww}. The first image of the M87* BH captured by the EHT qualitatively shows two distinct regions: a central dark region and a surrounding annular bright region. The bright region is formed by the photons reaching us from outside the horizon of the BH. The photon geodesics from the central region can't escape to the spatial infinity, rendering that region dark to us. The angular diameter of the dark region of the BH has been measured to be $42\pm3$ $\mu$as \cite{Akiyama:2019cqa,Akiyama:2019eap,Akiyama:2019fyp} with an axis ratio of $4/3$. The shadow is not entirely circular and the upper bound on the deviation of circularity is 10$\%$. The shadow of a rotating, uncharged black hole such as Kerr BH is highly circular irrespective of the inclination angle \cite{Johannsen:2010ru} assuming that the black hole can entirely be characterized by its mass and angular momentum. A deviation from circularity observed in the shadow of the M87* BH could be a sign of it being a non-Kerr BH. Several studies using various non-Kerr black holes with origin either in general relativity or in string theory have tried to explain the observed deviation \cite{Banerjee:2019nnj,Dokuchaev:2019jqq, Alexeyev:2020frp, Kumar:2020hgm}. 

Although the EHT collaboration has not released any polarimetric data, there have been several polarimetric observations of M87* in the past in multiple frequency bands\cite{Kuo:2014pqa,1990ApJ...362..449O,Algaba:2016tql,2001IAUS..205..136J,Zavala:2002zz}. The analysis in \cite{Kuo:2014pqa} concludes that due to the large uncertainties in the measurements it is difficult to distinguish between the Faraday rotation and the internal rotation of the polarization angle. Axions as a source of the rotation of the polarization plane of light has been studied in \cite{Chen:2019fsq}

In this work, we explore the possibility of the M87* being a Kerr-Sen black hole (KSBH) \cite{Sen:1992ua}. The KSBH emerges as a rotating charged black hole solution to the low energy 4-dimensional effective action of the heterotic string theory involving the dilaton, a $U(1)$ gauge field and an axion appearing through the Kalb-Ramond 3-form tensor. The Kalb-Ramond field is made up of an axion field and the Chern-Simons term for the gauge field assuming weak gravity in the region far away from the space-time singularity of the KSBH. Due to this, axions become a natural hair of the black hole instead of being a part of the cloud in the accretion disk. Circularly polarized photons coming from behind the black hole or those in the circular orbits around the black hole outside the horizon can interact with axions leading to axion-photon oscillation and birefringence such that their polarization angle gets rotated  \cite{Wilczek:1987mv,Carroll:1989vb,Carroll:1991zs,Harari:1992ea}. This rotation of the polarization (frequency independent ``Faraday rotation'') is an intrinsic feature of the black hole, distinguishing it from the usual Faraday rotation which is dependent on the wavelength of the photon. There have been several studies on the features of a KSBH including lensing \cite{Gyulchev:2006zg,Uniyal:2018ngj}, null geodesics and photon capture to produce a shadow \cite{Hioki:2008zw,Uniyal:2017yll} and comparison with other rotating charged black holes \cite{Campbell:1992hc,Hioki:2008zw}. We calculate the features of the shadow cast by a KSBH and also study the rotation of the polarization of circularly polarized light in the vicinity of such a black hole. An observation of a shadow and rotation of circularly polarized light consistent with KSBH could provide an indirect affirmation for string theory. To the best of our knowledge, there has been no other simultaneous study of the BH shadow and axion-photon interaction in the context of a KSBH.

The 4-dimensional heterotic string action has terms that are proportional to the inverse string tension $\alpha'$. Identifying the $U(1)$ gauge field with the photon allows us to get a relation between $\alpha'$ and the electric charge $e$ which is the $U(1)$ gauge coupling. Using this for calculation of the rotation of polarization of circularly polarized light from observations gives us the charge that can be carried by the KSBH which can then be used to calculate the deviation from circularity of a black hole image. The rotation by a constant angle of the polarization for all frequencies together with the deviation from circularity of the black hole image can be a signature of a KSBH.

This paper is organized in the following manner. In Sec. \ref{KSBH}, a brief introduction to the KSBH is given. We calculate the trajectories of the null geodesics ending in unstable circular orbits to get the shadow of the KSBH in Sec. \ref{image}. In Sec. \ref{Pol}, we calculate the rotation of the polarization of circularly polarized photons around the KSBH due to its axion hair. Finally we conclude in Sec. \ref{resdis} with discussions on the possible consequences of the results.

\section{Kerr-Sen Black Hole}\label{KSBH}
\begin{figure}[tbh!]
\begin{center}
\includegraphics[width=3.4in,height=3.4in,angle=0]{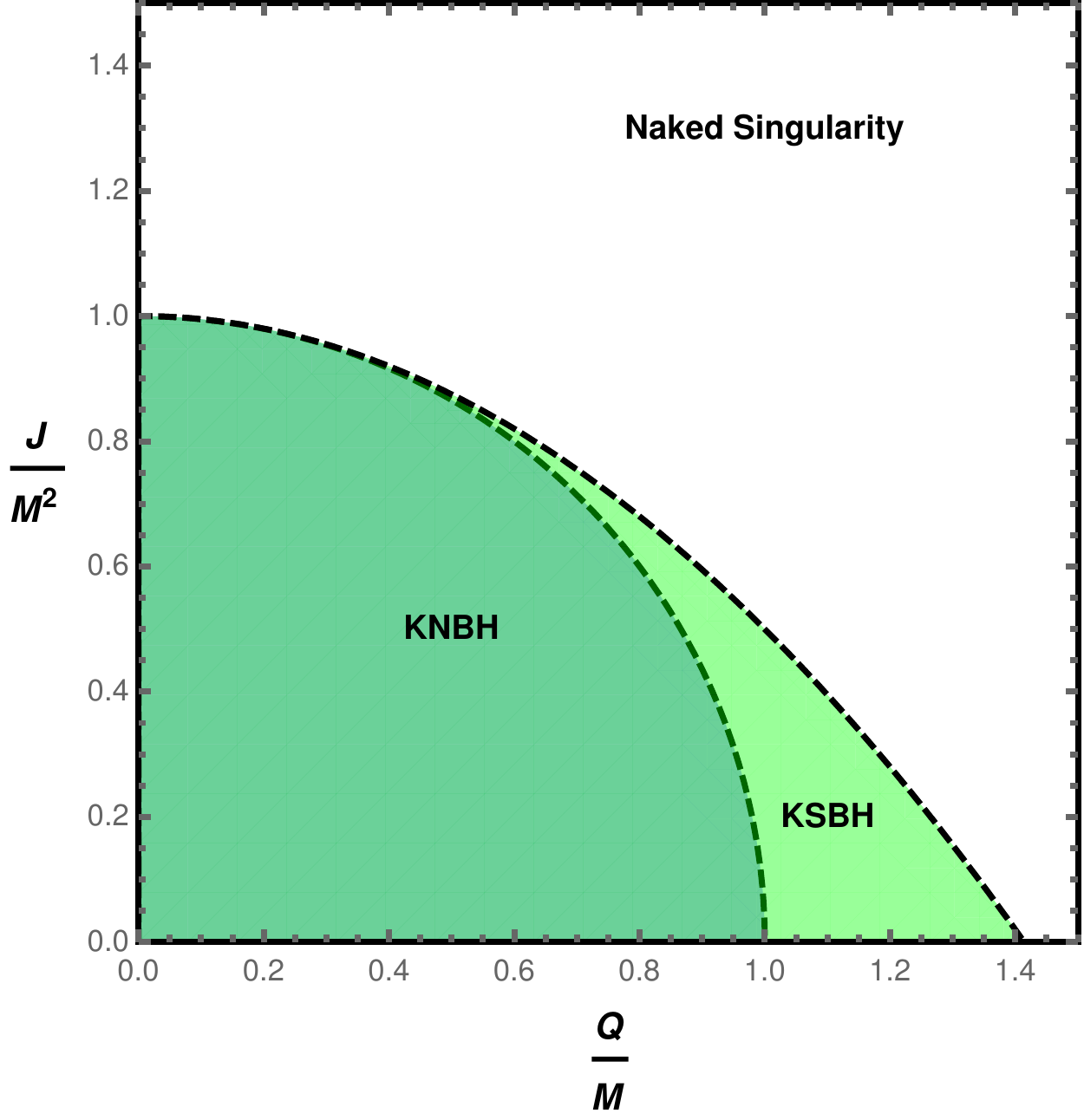}~~
\caption{The dashed line shows the region where $r_-=r_+$. This is the extremal value where horizons can exist. The shaded region is where the black hole is bound by horizons while the unshaded region depicts a naked singularity marked by an absence of horizons. It's also seen that the KSBH can carry more charge than the KNBH.}
\label{fig:horreg}
\end{center}
\end{figure}

In \cite{Sen:1992ua}, Sen constructed a rotating charged black hole solution by applying certain transformations \cite{Ferrara:1976iq,Cremmer:1977zt,Cremmer:1977tt,Cremmer:1977tc,Cremmer:1979up,deRoo:1985np,Castellani:1985ka,Castellani:1985wk,Cecotti:1988zz,Veneziano:1991ek,Duff:1989tf,Meissner:1991zj,Meissner:1991ge,Gasperini:1991qy,Sen:1991zi,Horne:1991cn} on the Kerr black hole solution \cite{Kerr:1963ud} starting from the string theory effective action in 4 dimensions. The form of the action used in \cite{Sen:1992ua} and written in the string frame is
\bea \label{eq:SenString}
S= \int d^4 x \sqrt{\mathcal{G}} e^{-\tilde{\phi}}
 \left(\tilde{R}^\prime -\frac{1}{12} \tilde{H}^{\prime 2}-
 \partial_{\mu}^{\prime}\tilde{\phi} \partial^{\mu \prime}
 \tilde{\phi}-\frac{\tilde{F}^{\prime 2}}{8} \right),
\eea
where $\mathcal{G}_{\mu\nu}$ is the string frame metric and the prime on various quantities denotes the string frame equivalents of the Ricci scalar $\tilde{R}$, the Kalb-Ramond 3-form $\tilde{H}_{\mu\nu\lambda}$, the gauge field $\tilde{F}_{\mu\nu}$ and $\phi$ is the dilatonic field used to make the conformal transformation from the Einstein frame to the string frame. This is a low energy effective string action written upto $\mathcal{O}(\alpha^\prime)$ where $\alpha^\prime$ is the inverse string tension.
The low energy effective heterotic string action in four dimensions upto $\mathcal{O}(\alpha^\prime)$ as given in \cite{Campbell:1992hc} is
\begin{eqnarray}\label{eq:campbellaction}
S=\int d^4x \sqrt{g}\left( \frac{R}{2\kappa^2}-6\,e^{-2\sqrt{2}\kappa\phi}H^{\mu\nu\lambda}H_{\mu\nu\lambda}-\frac{1}{2}\partial_\mu\phi\partial^\mu\phi -\frac{\alpha'}{16\kappa^2} e^{-\sqrt{2}\kappa\phi}F_{\mu\nu}F^{\mu\nu}   \right), 
\end{eqnarray}
 where  $H_{\mu\nu\lambda}$ is the Kalb-Ramond 3-form, $R$ is the Ricci scalar and $F_{\mu\nu}=\partial_\mu A_\nu-\partial_\nu A_\mu$ is the gauge field all in the Einstein frame while $\phi$ is the dilatonic field and $\kappa^2=M_{Pl}^{-2}$.
 We scale the action in Eq. (\ref{eq:campbellaction}) by $2\kappa^2$ and absorb the constants $\kappa$ and $\alpha'$ in the fields by the following redefinitions
\begin{eqnarray}
\tilde{H}_{\mu\nu\lambda}=12\kappa H_{\mu\nu\lambda},\nonumber \\
\tilde{\phi}=\sqrt{2}\kappa\phi,\\
\tilde{F}_{\mu\nu}=\sqrt{\alpha'}F_{\mu\nu} \nonumber,
\end{eqnarray}
In terms of the new fields $\tilde{H}_{\mu\nu\lambda},\;\tilde{F}_{\mu\nu}$ and $\tilde{\phi}$, the action in Eq. (\ref{eq:campbellaction}) becomes
\begin{eqnarray}\label{eq:SenEinstein}
S=\int d^4 x \sqrt{g} \left(R - \frac{1}{12}e^{-2\tilde\phi}\tilde{H}^2- \partial_\mu\tilde{\phi}\,\partial^\mu\tilde{\phi} -\frac{1}{8}e^{-\tilde\phi}\tilde{F}^2  \right).
 \end{eqnarray}
The action in Eq. (\ref{eq:SenEinstein}) is also the exact Einstein frame equivalent of the Sen action given in Eq. (\ref{eq:SenString}) obtained by transforming the string frame metric, $\mathcal{G}_{\mu\nu}$, to the Einstein frame metric, $g_{\mu\nu}$ by
\begin{equation}
\mathcal{G}_{\mu\nu}=e^{\tilde{\phi}}g_{\mu\nu}.
\end{equation}
Since we are interested in regions far from a space-time singularity, we can ignore terms with more than two derivatives of the metric in the action. This allows us to write the Kalb-Ramond 3-form as just a fully antisymmetrized derivative of an axion field and the gauge field Chern-Simons terms in the following manner
\begin{equation}
H_{\mu\nu\lambda}=\frac{e^{\sqrt{2}\kappa\phi}}{6\sqrt{2}}\epsilon_{\alpha\mu\nu\lambda}\partial^\alpha \chi-\frac{\alpha'}{32\kappa\sqrt{g}}A_{[\mu}F_{\nu\lambda]},
\label{eq:3form}
\end{equation}
where $\chi$ is the axion field and the square brackets around the indices denote a cyclic sum over the indices.
%
%\begin{equation}\label{eq:axionDef}
%\partial_\mu \chi = \frac{1}{6\sqrt{2}}e^{\sqrt{2}\kappa\phi}\partial_\mu b
%\end{equation}
In this work, we limit to the $U(1)$ gauge field. Substituting $H_{\mu\nu\lambda}$ in Eq. (\ref{eq:campbellaction}) in favour of the axion and the gauge fields and working in the region where the classical value of the dilaton, $\phi$, is small (anticipating $\phi \propto \alpha'$) such that $e^\phi\simeq 1$, we get
\begin{dmath}
S=\int d^4 x \sqrt{g}\left[\frac{R}{2\kappa^2}-\frac{1}{2}\partial_\mu\phi\partial^\mu\phi-\frac{1}{2}\partial_\mu\, \chi\;\partial^\mu\, \chi-\frac{\alpha'}{16\kappa^2}F_{\mu\nu}F^{\mu\nu}-\frac{\alpha'}{8\kappa}\frac{6\sqrt{2}}{4!}\frac{\epsilon^{\mu\nu\lambda\sigma}}{\sqrt{g}}\chi\;F_{\mu\nu}F_{\lambda\sigma}\right].
\label{eq:act2}
\end{dmath}
The equations of motion for $\chi$, $\phi$ and $F^{\mu \nu}$ respectively are 
\begin{eqnarray}\label{eq:fieldEom}
\square \chi &=& \frac{\alpha'}{8\kappa}\frac{6\sqrt{2}}{4!}F_{\mu\nu} (^\star F)^{\mu\nu}, \\
\square \phi &=& -\frac{\sqrt{2}\alpha'}{16\kappa}F_{\mu\nu}F^{\mu\nu},\\
\nabla_{\mu}F^{\mu \nu}&=&-\frac{\kappa}{\sqrt{2}}\left(\partial_{\alpha} \chi \right)(^
\star{F})^{\alpha \nu},
\end{eqnarray}
where $(^\star F)^{\mu\nu}=\frac{\epsilon^{\mu\nu\lambda\sigma}}{\sqrt{g}}F_{\lambda\sigma}$ is the dual of $F_{\mu\nu}$, $\square$ is the d'Alembertian operator: $\square = \nabla_\mu\nabla^\mu$ and $\nabla_\mu$ is the covariant derivative operator.
The action given in Eq. (\ref{eq:SenEinstein}) admits the following metric (named the Sen metric) as one of the inequivalent solutions obtained by the twisting procedures shown in \cite{Ferrara:1976iq,Cremmer:1977zt,Cremmer:1977tt,Cremmer:1977tc,Cremmer:1979up,deRoo:1985np,Castellani:1985ka,Castellani:1985wk,Cecotti:1988zz,Veneziano:1991ek,Duff:1989tf,Meissner:1991zj,Meissner:1991ge,Gasperini:1991qy,Sen:1991zi,Horne:1991cn}:
\begin{eqnarray}
\nonumber ds^{ 2}&=& -{(r^2 +a^2\cos^2\theta -2\mu r)
(r^2+a^2\cos^2\theta) \over (r^2+a^2\cos^2\theta
+2 \mu r\sinh^2{\alpha\over 2})^2} dt^2
\nonumber \\ && +{r^2 +a^2\cos^2\theta\over r^2
+a^2 -2\mu r} dr^2 + (r^2 +a^2\cos^2\theta) d\theta^2
\nonumber \\ &&+\{(r^2+a^2)(r^2+a^2\cos^2\theta) +2 \mu r a^2\sin^2\theta +4\mu r
(r^2+a^2) \sinh^2{\alpha\over 2}
\nonumber \\ &&+ 4\mu^2 r^2\sinh^4{\alpha\over 2}\}
 \times {(r^2+a^2\cos^2\theta)\sin^2\theta\over (r^2
+a^2\cos^2\theta +2\mu r\sinh^2{\alpha\over 2})^2} d\phi^2
\nonumber \\ &&- {4\mu r a\cosh^2{\alpha\over 2} (r^2+a^2\cos^2\theta)\sin^2\theta
\over (r^2
+a^2\cos^2\theta +2\mu r\sinh^2{\alpha\over 2})^2} dt d\phi,
\label{eq:kerrsen}
\end{eqnarray}
and the field solutions are
\begin{eqnarray}
\chi&=&- Q^2  \frac{\alpha^{\prime}}{\kappa} \frac{a}{GM} \frac{\cos \theta}{r^2+a^2 \cos \theta}, \\
A_{t}&=&-\frac{1}{\sqrt{\alpha'}} \left(\frac{2 \mu r a \sinh \alpha \sin^2 \theta}{r^2 + a^2 \cos^2 \theta + 2 \mu r \sinh^2 \frac{\alpha}{2}} \right), \\
A_{\phi}&=&\frac{1}{\sqrt{\alpha'}} \left(\frac{2 \mu r \sinh \alpha}{r^2 +a^2 \cos^2 \theta +2 \mu r \sinh^2 \frac{\alpha}{2}} \right),
\label{eq:axionsol}
\end{eqnarray}
where the constants $\mu$, $\alpha$ and $a$ give the physical mass $M$, charge $Q$ and angular momentum $J$ through multipole expansions of $g_{tt}$, $A_t$ and $g_{t\phi}$ respectively in powers of $1/r$,
\begin{eqnarray}
GM=\frac{\mu}{2}\left( 1+ \cosh \alpha \right), \;\;\;\sqrt{\alpha'} Q=\frac{\mu}{\sqrt{2}}\sinh \alpha, \;\;\; J=\frac{a \mu}{2}\left( 1+\cosh \alpha \right).
\end{eqnarray}
The metric in Eq. (\ref{eq:kerrsen}) has a space-time singularity at $r=0$ bound by two horizons at $r_-$ and $r_+$ which encapsulate a rotating charged black hole, called the Kerr-Sen (KS) black hole. In terms of $M$, $Q$ and $J$, $r_\pm$ are
\begin{eqnarray}
r_{\pm}=GM-\frac{Q^2}{2M}\pm \sqrt{\left(GM-\frac{Q^2}{2M}\right)^2-\frac{J^2}{M^2}}.
\end{eqnarray}
The existence condition of the horizons ($r_\pm$ must be real and non-negative) gives a bound for the mass, charge and rotation of the black hole inside the horizon
\begin{eqnarray}
\left(GM-\frac{Q^2}{2M}\right)^2\geq \frac{J^2}{M^2},
\label{eq:Regular}
\end{eqnarray}
called horizon regularity condition which is shown in the Fig. \ref{fig:horreg} along with a comparison of the horizon regularity for the charged rotating Kerr-Newman black hole (KNBH) arising from general relativity. A feature that distinguishes the KS black hole from the KNBH is its increased intrinsic charge carrying capacity without falling into a naked singularity. 

\section{Black hole image and deviation from circularity}
\label{image}
Photons coming from a light source behind the black hole are either scattered away or fall into the black hole depending on their impact parameter from the black hole. The null geodesics which describe the circular photon orbits around the black hole are specially interesting because of their observational importance. These geodesics describe a capture region which can be understood as the shadow of the black hole cast at an observer at infinity in the impact parameter space \cite{Vries_1999,ZAKHAROV2005479,refId0}. This shadow can be observed in telescopes like the Event Horizon Telescope \cite{Akiyama:2019cqa,Akiyama:2019fyp,Akiyama:2019eap} to give information about the space-time behavior in a region of strong gravity near a black hole. Calculations in this section are done in geometric units with $G=c=1$. 
For convenience, we use the Boyer-Lindquist coordinates \cite{doi:10.1063/1.1705193} as follows
\begin{eqnarray}
	&&
	\Delta \equiv r(r+r_{0}) - 2 M r +a^2 \ ,
\;\;\;
	\rho^2 \equiv r(r+r_{0}) + a^2 \cos^2 \theta \ .
\end{eqnarray}
In terms of these new coordinates, the KS metric in Eq. (\ref{eq:kerrsen}) can be written as
\begin{eqnarray}
	&&
	ds^2\nonumber
	=
	-\left(
		1-\frac{2 M r}{\rho^2}
	\right) dt^2
	-
	\frac{4 M r a \sin ^2 \theta}{\rho^2} dtd\varphi
	+
	\frac{\rho^2}{\Delta } dr^2 
	+
	\rho^2 d\theta ^2 \\ && 
	+ 
	\left(
	r(r+r_{0})+a^2+\frac{2 M r a^2 \sin^2\theta }{ \rho^2 } \right) \sin^2\theta d\varphi ^2, 
\label{eq:metricBL}
\end{eqnarray}
with $r_{0}=Q^2/M$.

\subsection{Hamilton-Jacobi equation}
We use the formalism given in \cite{Carter:1968rr} to find the separate equations of motion. The Hamilton-Jacobi equation is given by
\begin{eqnarray}
2 \frac{\partial S}{\partial \lambda}=g^{\mu \nu}\frac{\partial S}{\partial x^{\mu}}\frac{\partial S}{\partial x^{\nu}},
\label{eq:HJeq}
\end{eqnarray}
where $S(\lambda, x^{\mu})$ is the Jacobian action and $\lambda$ is the affine parameter. Motion of a test particle in an axisymmetric space-time, such as Kerr-Sen space-time, has two conserved quantities, angular momentum $L_z$ with respect to the rotation axis of the black hole and energy $E$, associated with two Killing vectors $\partial_{\phi}$ and $\partial_{t}$ respectively. Since the Hamiltonian is also independent of the affine parameter $\lambda$, the solution for $S$ can be separated in terms of these to get
\begin{equation}
S=\frac{1}{2}m^2 \lambda-E t + L_z \phi +S_{r\theta}(r,\theta).
\end{equation}
$S_{r\theta}$ can be further separated into $S_r(r)$ and $S_\theta(\theta)$ as the metric in Eq. (\ref{eq:metricBL}) doesn't have any $g_{r\theta}$ term, to finally give \cite{Blaga:2001wt} 
\begin{eqnarray}
S(\lambda, x^{\mu})=\frac{1}{2}m^2 \lambda-E t + L_{z} \phi + S_{r}(r)+S_{\theta}(\theta),
\label{eq:jacsol}
\end{eqnarray} 
where $m$ is mass of the test particle. 

%The line element in eq. (\ref{eq:metricBL}) can be written in terms of the inverse metric as
%\begin{eqnarray}
%ds^2=\frac{\Sigma}{\rho^2 \Delta}\left(\frac{\partial}{\partial t}\right)^2-\frac{4 M r a}{\rho^2 \Delta}\left(\frac{\partial}{\partial t}\frac{\partial}{\partial \phi}\right) + \frac{\Delta}{\rho^2}\left(\frac{\partial}{\partial r}\right)^2+\frac{1}{\rho^2}\left(
%\frac{\partial}{\partial \theta}\right)^2+\frac{\Delta-a^2 \sin^2\theta}{\rho^2 \Delta \sin^2 \theta}\left(\frac{\partial}{\partial \phi}\right)^2,
%\label{eq:invmet}
%\end{eqnarray}
%where

Using Eq. (\ref{eq:jacsol}) in Eq. (\ref{eq:HJeq}) we get the separated radial and azimuthal parts of Hamilton-Jacobi equation \cite{Vries_1999,Blaga:2001wt}
\begin{eqnarray}
\Delta^2 \left(\frac{d S_{r}}{d r}\right)^2&=&\mathcal{R}(r)\Rightarrow S_r(r)=\int^{r} dr^{\prime} \frac{\sqrt{\mathcal{R}(r^{\prime})}}{\Delta}\label{eq:dSr},\\
\left(\frac{d S_{\theta}}{d \theta}\right)^2&=&\Theta(\theta)\Rightarrow S_\theta(\theta)=\int^{\theta} d \theta^{\prime} \sqrt{\Theta(\theta^{\prime})}\label{eq:dSazim},
\end{eqnarray}
with
\begin{eqnarray}
\mathcal{R}(r)&=&\left[-\left\lbrace r(r+r_0)+a^2\right\rbrace E+a L_z\right]^2-\Delta \left[\mathcal{Q}+(L_z-a E)^2+m^2 r(r+r_0)\right],\\
\Theta(\theta)&=&\mathcal{Q}-\left( L_{z}^{2} \csc^2 \theta- a^2 E^2 \right) \cos^2 \theta-m^2 a^2 \cos^2 \theta,
\end{eqnarray}
where $\mathcal{Q}$ is the constant of separation, also called the Carter constant \cite{Carter:1968rr}.
 
Using the definitions $p_{\mu} \equiv \partial S/\partial x^{\mu}$ and $\partial x^{\mu}/\partial \lambda \equiv g^{\mu \nu} p_{\nu}$ we find the following equations of motion for the radial and azimuthal motion of the particle
\begin{eqnarray}
\rho^4 \left( \frac{d r}{d \lambda}\right)^2&=&\mathcal{R}(r),\label{eq:rGeo}\\
\rho^4 \left( \frac{d \theta}{d \lambda}\right)^2&=&\Theta(\theta),\label{eq:thetaGeo}\\
%\rho^4 \left( \frac{d t}{d \lambda}\right)^2=Something\\
\rho^4 \left( \frac{d \phi}{d \lambda}\right)^2&=&\frac{1}{\Delta^2}\left[ 2 M a r E + L_z \csc^2 \theta \left( \Sigma-2 M r\right), \right]^2\label{eq:phiGeo},
\end{eqnarray}
where
\begin{eqnarray}
\Sigma\equiv\left[r(r+r_0)+a^2 \right]^2-a^2 \Delta \sin^2 \theta.
\end{eqnarray}
For a massless particle such as a photon coursing along null geodesics, $m=0$, and the potentials $\mathcal{R}(r)$ and $\Theta(\theta)$ are parametrized only by $E$, $L_z$ and the Carter's constant $\mathcal{Q}$. Scaling these potentials by $E^2$ such that $\mathcal{R}(r)=\overline{\mathcal{R}}(r)E^2$ and $\Theta(\theta)=\overline{\Theta}(\theta) E^2$ and defining
\begin{eqnarray}
\xi \equiv \frac{L_z}{E}, \;\;\; \eta \equiv \frac{\mathcal{Q}}{E^2},
\end{eqnarray}
we get the following form for the $E^2$ scaled potentials $\overline{\mathcal{R}}$ and $\overline{\Theta}$
\begin{eqnarray}
\overline{\mathcal{R}}(r)&=&\left[-\left\lbrace r(r+r_0)+a^2\right\rbrace +a \xi \right]^2-\Delta \left[\eta+(a-\xi)^2\right]\label{eq:Req},\\
\overline{\Theta}(\theta)&=&\eta-\left( \xi^2 \csc^2 \theta- a^2 \right) \cos^2 \theta,
\label{eq:thetaeq}
\end{eqnarray}

From Eqs. (\ref{eq:Req}) and (\ref{eq:thetaeq}) it's clear that the motion of a photon on a null geodesic is parametrized only by two parameters $\xi$ and $\eta$ \cite{Gyulchev:2006zg}.

\subsection{Circular photon orbits}
Circular photon orbits are characterized by the vanishing of the radial potential and its derivative at $r_{cpo}$,
\begin{eqnarray}
\overline{\mathcal{R}}(r_{cpo})&=& 0, \nonumber\\ 
\partial_{r}\overline{\mathcal{R}}(r_{cpo})&=& 0,
\label{eq:cpocond}
\end{eqnarray}
where $r_{cpo}$ is the radius of the photon orbit. The conditions in Eq. (\ref{eq:cpocond}) can be used to solve for $\xi$ and $\eta$ in Eq. (\ref{eq:Req}). There are two pairs of solutions
\begin{eqnarray}
\xi=\frac{r(r+r_0)+a^2}{a}, \;\;\; \eta=-\frac{r^2(r+r_0)^2}{a^2},
\label{eq:pair1}
\end{eqnarray}
and,
\begin{eqnarray}\label{eq:pair2}
\xi &=& \frac{a^2 \left(-2 r_{\text{cpo}}-2 M-r_0\right)+6 M r_{\text{cpo}}^2+2 M r_0 r_{\text{cpo}}-2 r_{\text{cpo}}^3-3 r_0 r_{\text{cpo}}^2-r_0^2 r_{\text{cpo}}}{a \left(r_{\text{cpo}}-M+\frac{r_0}{2}\right)},\nonumber\\
\eta &=& \frac{a^2 r_{\text{cpo}}^2 \left(16 M r_{\text{cpo}}+8 M r_0\right)-r_{\text{cpo}}^2 \left(\left(r_{\text{cpo}}+r_0\right) \left(2
   r_{\text{cpo}}+r_0\right)-2 M \left(3 r_{\text{cpo}}+r_0\right)\right){}^2}{a^2 \left(2 r_{\text{cpo}}-2 M+r_0\right)^2},
\end{eqnarray}
for the solution pair in Eq. (\ref{eq:pair1}) the azimuthal potential $\Theta$ becomes negative which isn't allowed (see Eq. (\ref{eq:dSazim})). On the other hand, the solution pair in Eq. (\ref{eq:pair2}) does give a consistent solution for the circular photon orbits.  

Black hole image observations rely on photons that can reach the observer. Therefore, we need to look at unstable circular orbits of photons orbiting outside the exterior horizon of the KSBH, $r_{cpo}>r_+$. These photons reach the observer after a finite number of rotations around the black hole. If the distance between the black hole and the observer is $l_0$ and the angle of inclination (also called the viewing angle) of the black hole shadow  is $\theta$, then the photon has impact parameters (also known as the celestial coordinates) given by \cite{Young:1976zz,Gooding:2008tf} defined by

\begin{eqnarray}
\alpha=\lim_{l_0\to \infty}\left(-l_{0}^{2}\sin \theta \frac{d \phi}{d r} \right), \beta=\lim_{l_0\to \infty}\left(l_{0}^{2} \frac{d \theta}{d r} \right).
\label{eq:celcodef}
\end{eqnarray}
Using Eqs. (\ref{eq:rGeo}-\ref{eq:phiGeo}) in Eq. (\ref{eq:celcodef}) we get
\begin{eqnarray}
\alpha=-\xi\csc\theta, \;\;\; \beta=\pm \sqrt{\eta+a^2\cos^2 \theta-\xi^2 \cot^2 \theta}.
\label{eq:celfin}
\end{eqnarray}

\begin{figure}[h!]
\begin{center}
\includegraphics[width=4.9in,height=3.2in,angle=0]{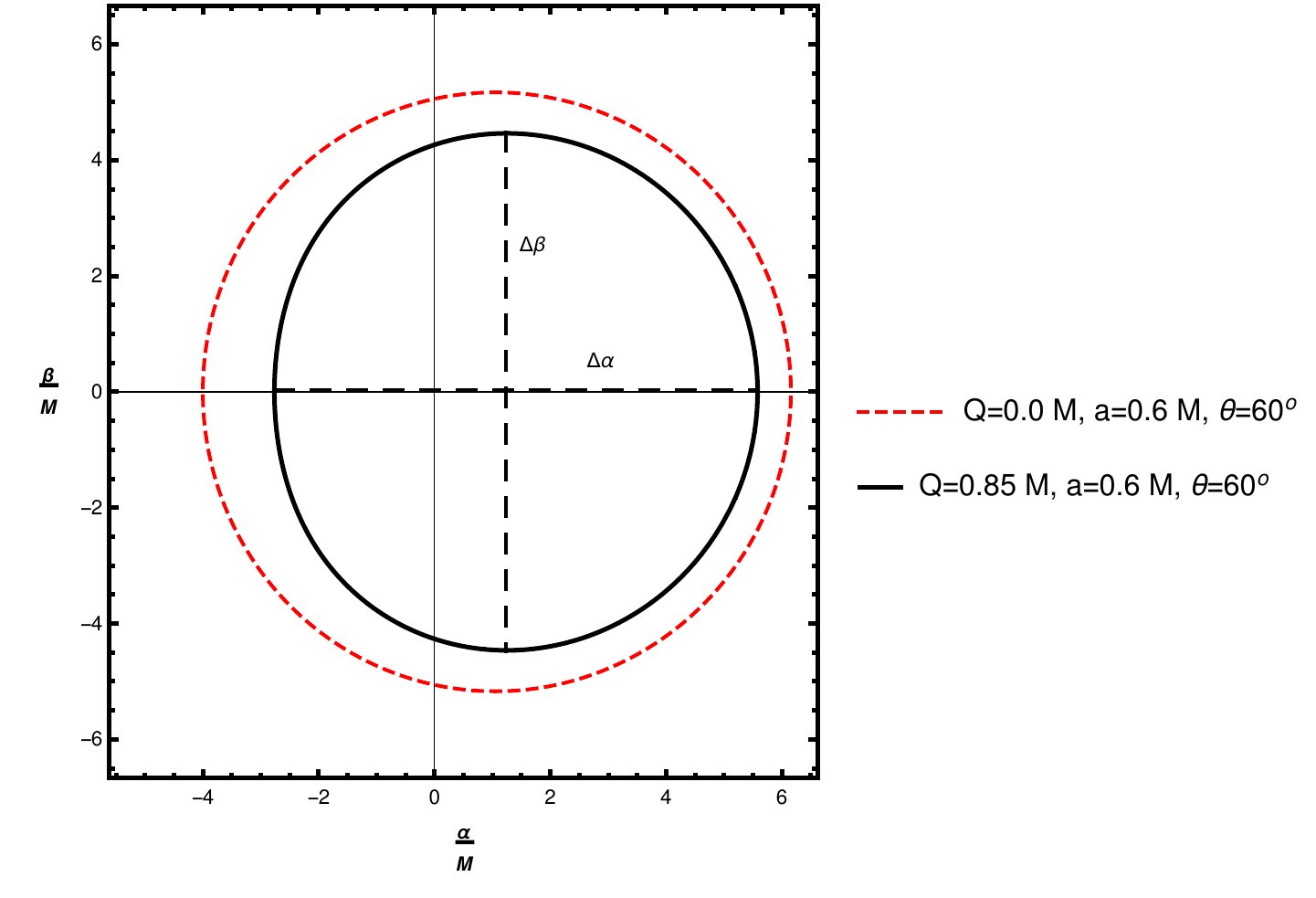}~~
\caption{For fixed inclination angle $\theta$ and spin $a$ the photon capture region gets smaller and deviates from circularity as the charge increases. Here we show the photon capture region for $\theta=60^o$ and $a=0.6 M$  for $Q=0$ (red, dashed)  and $Q=0.85 M$ (black, solid).}\label{fig:imagetemp}
\end{center}
\end{figure}

A representative image/shadow for the KSBH for the mass normalized celestial coordinates $\alpha/M$ and $\beta/M$ is shown the Fig. \ref{fig:imagetemp}.

\subsection{Deviation of the black hole shadow from circularity}
As can be seen in Fig. \ref{fig:imagetemp}, the image of the black hole is not an exact circle for non-zero values of the black hole charge, angular momentum and its inclination angle with respect to the observer. The amount of deviation from circularity is a measurable quantity and can give insight into the values of certain black hole hairs e.g. its charge and angular momentum. The image of the black hole is symmetric under reflection around $\beta=0$ which is the $\alpha$-axis. This tells us that the geometric center of this diameter is at $\beta=0$. However, the geometric center under reflections around the $\beta$ axis is shifted from $\alpha=0$ and it's not symmetric as well. The geometric center of the image on the $\alpha$ axis is obtained by taking its mean $\alpha_C = \int \alpha dA/\int dA$ where $dA$ is the area element of the image and fixing $\beta=0=\beta_C$.

One can now define the angle $\gamma$ that a point $(\alpha, \beta)$ on the boundary of the image subtends on the $\alpha$ axis at the geometric center, $\left ( \alpha_C, 0 \right )$. If $\ell(\gamma)$ is the distance between the point $(\alpha,\beta)$ and $(\alpha_C,0)$, the average radius $\bar{R}$ of the image is given by \cite{Bambi:2019tjh}
\begin{eqnarray}
	 R_{avg}^{2} &\equiv & \frac{1}{2\pi}\int_{0}^{2\pi} d\gamma \, \ell^2(\gamma)\,\;\;\\ \textrm{where},\,\;\; \nonumber 
	\ell(\gamma) &\equiv & \sqrt{ \left (\alpha(\gamma)-\alpha_C \right )^2 + \beta(\gamma)^2}\,. 
	\label{eq:r}
\end{eqnarray}
Finally, the deviation from circularity, $\Delta C$, is defined as the RMS difference of the distance between any point on the boundary and the center from the average radius of the shadow $R_{\text avg}$ \cite{Bambi:2019tjh},
\begin{eqnarray}
	\Delta C \equiv \frac{1}{R_{avg}}\sqrt{\frac{1}{2\pi}\int_0^{2\pi} d\gamma \left (\ell(\gamma) - R_{avg} \right )^2 }\,.
	\label{eq:devcircularity}
\end{eqnarray}
Another parameter that can be defined to quantify non-circularity of the photon shadow is the ratio of the diameters along the two axes \cite{Tsupko:2017rdo}
\begin{eqnarray}
D=\frac{\Delta \beta}{\Delta \alpha}.
\end{eqnarray}

For an inclination angle of $17^o$, The EHT collaboration observed the angular diameter of the shadow of M87* BH to be $(42\pm3) \mu \rm  a s$ \cite{Akiyama:2019cqa,Akiyama:2019fyp,Akiyama:2019eap} with the axis ratio 4/3 which translates to an upper bound of 10 percent on $\Delta C$. Since $\Delta C$ is a function of $a$ and $Q/M$, this bound can be used to put constraints on the spin and charge of the black hole. 

\begin{figure}[h!]
\begin{center}
\includegraphics[width=3in,height=3in,angle=0]{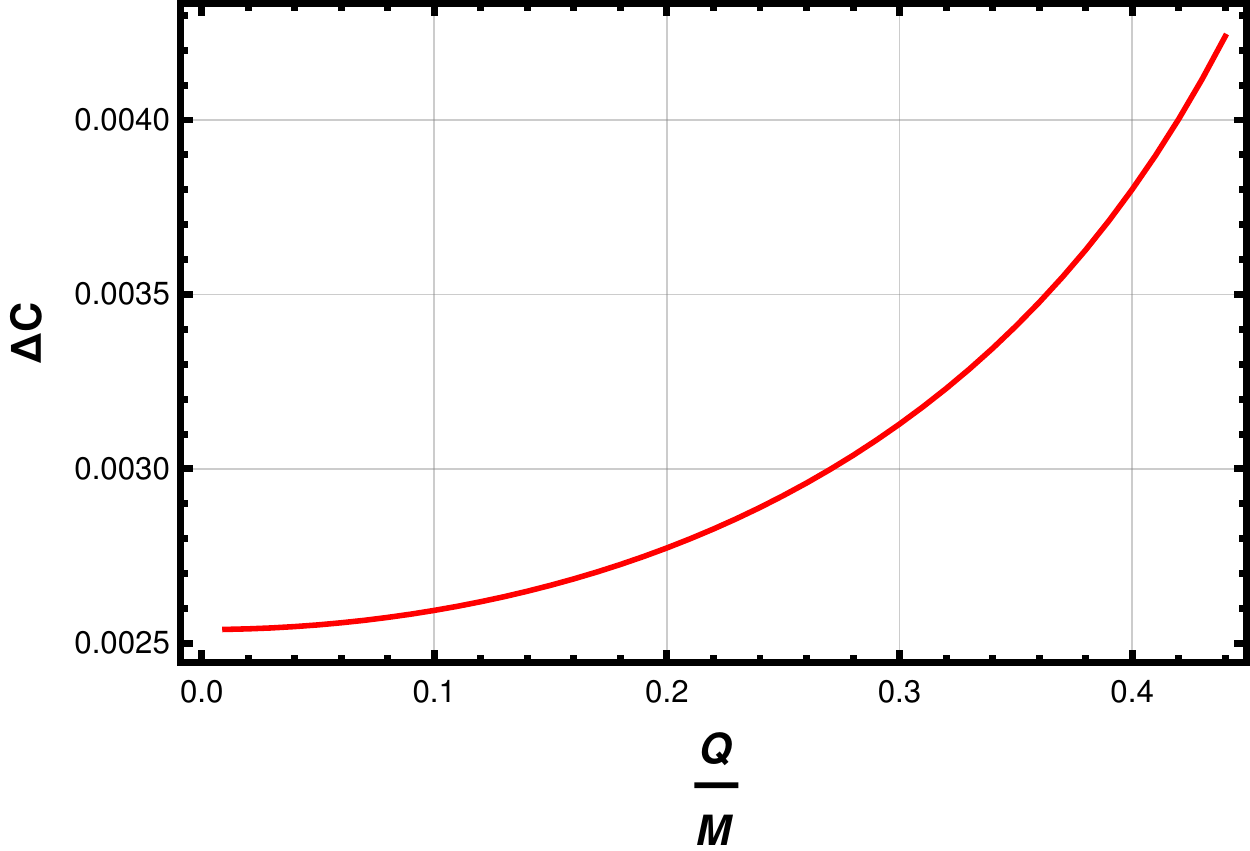}~~
\includegraphics[width=3in,height=3in,angle=0]{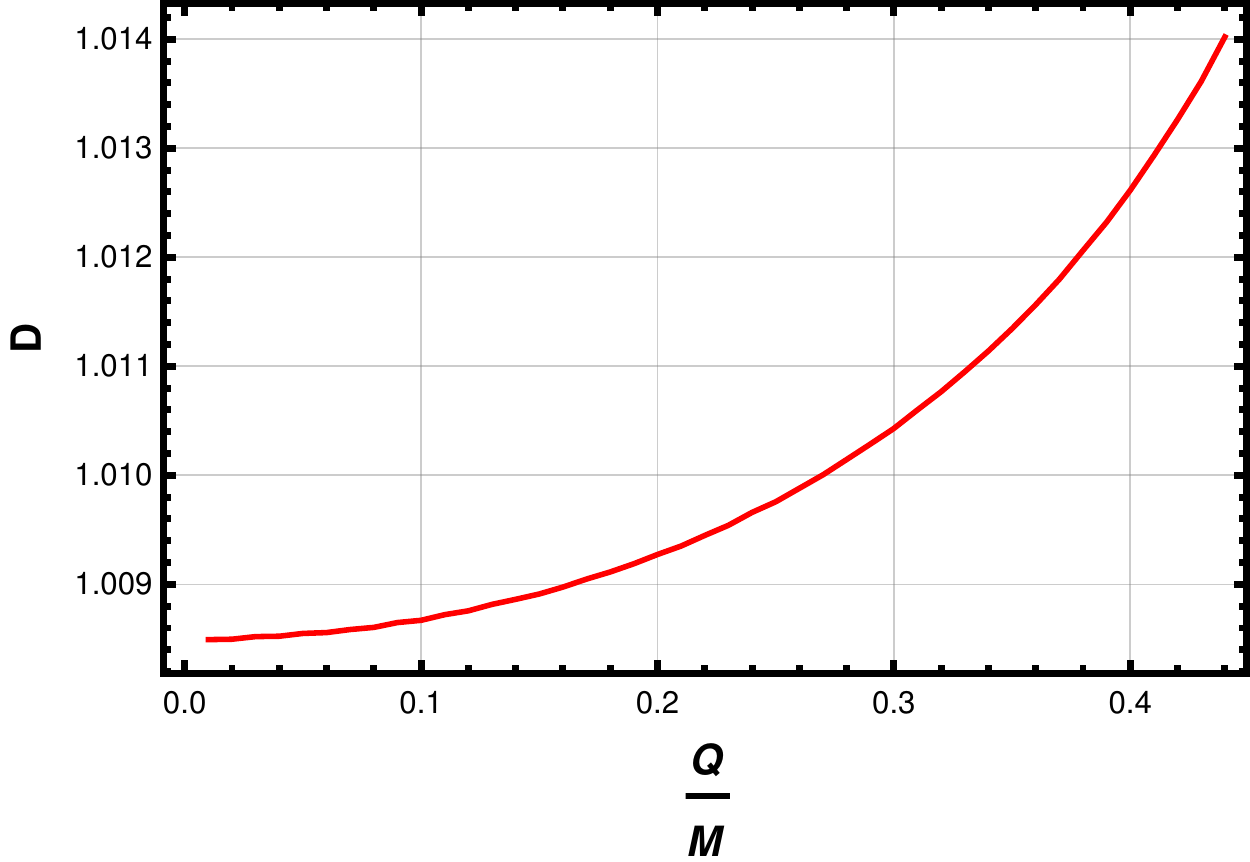}~~
\caption{The deviation from circularity, $\Delta C$, and the axis-ratio, $D$, as a function of $Q$ for $a=0.9 M$ and $\theta=17^0$.}\label{fig:devvscharge}
\end{center}
\end{figure}

The variation of $\Delta C$ and $D$ with charge for $\theta=17^o$ and $a=0.9 M$ is shown in Fig. \ref{fig:devvscharge}. As can be seen that $\Delta C$ and $D$ do not reach the upper bound reported by the EHT collaboration which implies that with the accuracy of the EHT observation we cannot put any constraints on spin or charge of the KSBH.

\section{Polarisation of light by axions}\label{Pol}
 It is well known that the axion-photon coupling gives rise to a birefringence for opposite-helicity photons propagating through a non-stationary or non-uniform axion field. This birefringence effect shows up in the rotation of the plane of polarization of the circularly polarized light \cite{Wilczek:1987mv,Carroll:1989vb,Carroll:1991zs,Harari:1992ea}.

The Lagrangian to study the polarization of light signal due to the axion cloud of the KSBH is (see Eq. (\ref{eq:act2}))
\begin{equation}\label{eq:axionLag}
{\cal L} \supset \frac{\alpha^{\prime}}{16 \kappa^2}  F_{\mu \nu}F^{\mu \nu}  + \frac{\alpha^{\prime}}{8 \kappa} \frac{6 \sqrt{2}}{4!}\chi F_{\mu \nu}\tilde{F}^{\mu \nu}.
\end{equation}
Assuming that the variation of the metric is much slower compared to that of the gauge field over space and time, the equation of motion of $F_{\mu\nu}$ reduces to:
\begin{eqnarray}
\partial_{\mu}F^{\mu \nu}=-\frac{\kappa}{\sqrt{2}}\left(\partial_{\alpha} \chi \right)\tilde{F}^{\alpha \nu}.\label{eq:FEqn}
\end{eqnarray}

Using the Lorentz gauge, Eq. (\ref{eq:FEqn}) reduces to the following equation for the gauge field $A_\mu$, 
\begin{eqnarray}
\square A^{\nu}=\frac{\kappa}{\sqrt{2}}\left(\partial_{\alpha} \chi \right)\tilde{F}^{\alpha \nu}.
\end{eqnarray}

We fix the axes of propagation such that the plane wave solution for $A_\mu$ propagates along the $z$-direction i.e.
\begin{equation}
A^\mu=A^\mu_0 e^{i\omega t- i k z},
\end{equation}
where the frequency, $\omega$, and the wave-number, $k$, obey the dispersion relation
\begin{eqnarray}
\omega=k \pm \frac{i \kappa}{2 \sqrt{2}}\left(\partial_{z} \chi\right).\label{eq:disp}
\end{eqnarray}

We fix $A_0=A_3=0$ for circularly polarized light and define $A_\pm=A_1\pm i A_2$ to finally obtain
\begin{eqnarray}
\begin{aligned} A _ { \pm } ( t , z ) & = A _ { \pm } \left( t ^ { \prime } , z ^ { \prime } \right) \exp \left[ - i \omega _ { \gamma } \left( t - t ^ { \prime } \right) + i \omega _ { \gamma } \left( z - z ^ { \prime } \right) \right.\\ &\left. \pm i \frac{\kappa}{2 \sqrt{2}}\left(\partial_{z} \chi\right) \right], \end{aligned}
\end{eqnarray}
where $\pm$ denotes the left and right circularly polarized light respectively. It's evident that the axion-photon interaction term produces a rotation of the electromagnetic wave by the angle $\Delta \Theta$ given by
\begin{eqnarray}
\nonumber
\Delta \Theta &=&\int dz \frac{\kappa}{2 \sqrt{2}}\left(\partial_{z} \chi \right)\\ \nonumber
&=& \frac{\kappa}{2 \sqrt{2}} \left(\chi(z_{obs})-\chi(z_{emit})\right)\\
&=&-\frac{\kappa}{2 \sqrt{2}} \chi(z_{emit}).
\label{eq:polmeas}
\end{eqnarray}
Substituting the axion solution from Eq. (\ref{eq:axionsol}) in Eq. (\ref{eq:polmeas}), the angle of rotation turns out to be
\begin{equation}
\Delta \Theta = \frac{ \alpha^{\prime}}{32} \frac{Q^2 a}{GM} \frac{\cos \theta}{r^{2} + a^2 \cos^2\theta}.
\label{eq:polalpha}
\end{equation}

The EHT results from the observation of M87* black hole do not include the polarimetric data at present. In \cite{Kuo:2014pqa}, M87* BH is observed in four frequency bands and it is assumed in the analysis that the position polarization angle follows the $\lambda^2$ law (Faraday rotation). However, the large uncertainties in the data (see Table \ref{tab:pol}, data from \cite{Kuo:2014pqa}) indicate that the polarization is not necessarily due to the Faraday rotation. This can be the polarization arising from the photon-axion interaction since it is independent of the wavelength of the photon. 
\begin{table}[h!]
\begin{center}
\begin{tabular}{ccc}
\hline
Frequency ($\nu$ GHz) & Polarization Angle ($\Delta \Theta$) &  \\ \hline
230.3 & $30.7\pm 2.2$ &  \\
232.3 & $31.6\pm 2.0$ &  \\
220.4 & $31.4\pm 1.9$ &  \\
218.4 & $27.3\pm 2.0$ &  \\ \hline
\end{tabular}

\end{center}
\caption{The observed polarization angle for different frequencies taken from \cite{Kuo:2014pqa}. The observations suggest that the polarization angle is independent of the frequency and can be due to some intrinsic feature of the black hole.}\label{tab:pol}
\end{table}

\begin{figure}[h!]
\begin{center}
\includegraphics[width=4in,height=4in,angle=0]{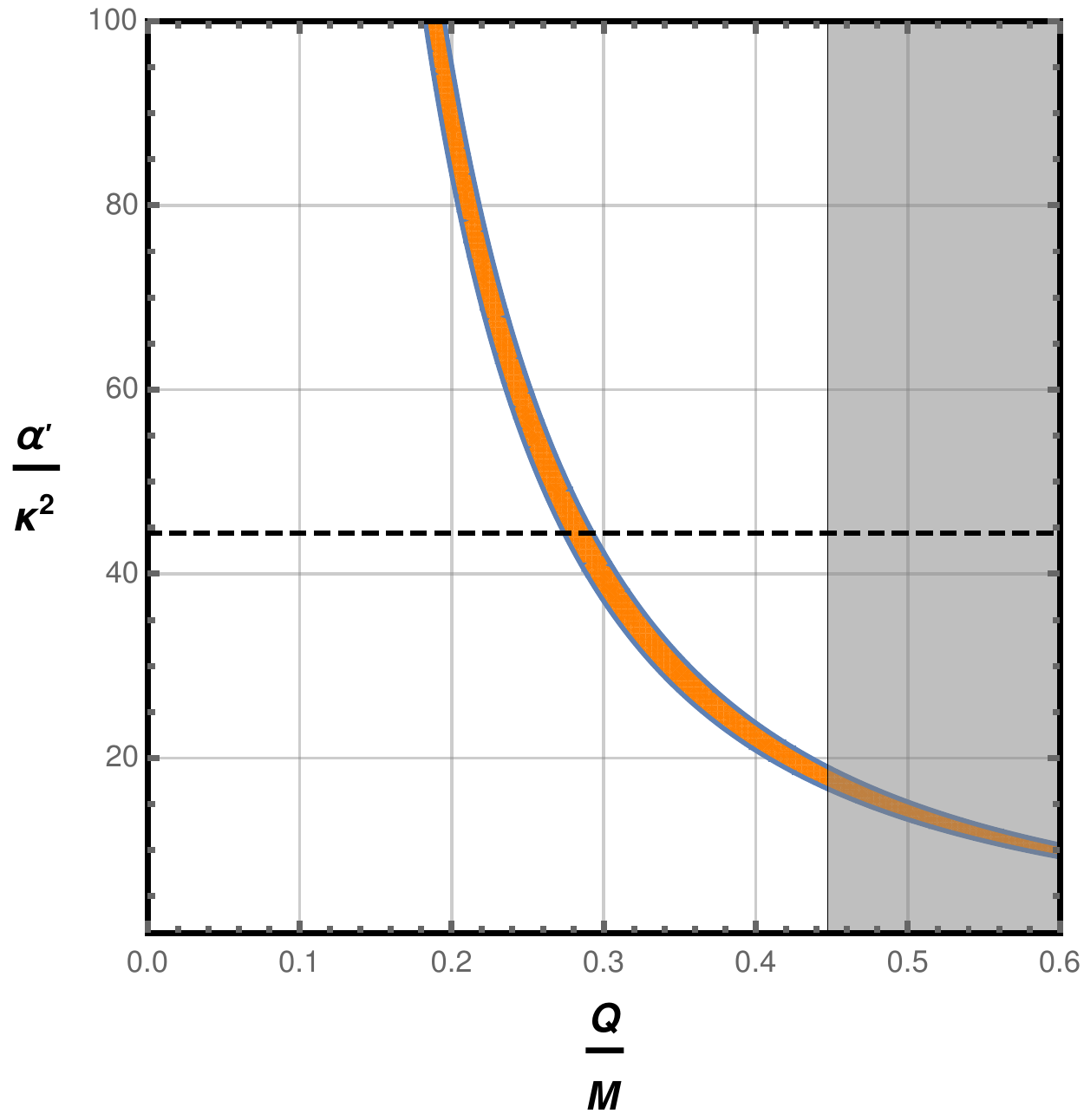}~~
\caption{The variation of $\alpha'/\kappa^2$ with the charge in the KSBH. The horizontal line marks the value of $\alpha'$ corresponding to the electromagnetic coupling $e$ and the values of $Q/M$ in the shaded region on the right are excluded for spin $a=0.9 M$ from the horizon regularity condition.}\label{fig:boundonalpha}
\end{center}
\end{figure}

Eq.(\ref{eq:polalpha}) shows that the polarization of light by the axion hair of the KSBH is dependent on the inverse string tension $\alpha'$. Assuming that the axion hair is the only source of polarization, we show the behaviour of $\alpha'$ with the charge of the KSBH in Fig. \ref{fig:boundonalpha}. We use the average, $30.25\pm 2.02$, of the four polarization angles in Table \ref{tab:pol} in our estimation of $\alpha^\prime$. The $U(1)$ gauge kinetic term in Eq. (\ref{eq:axionLag}) comes with a pre-factor of $\alpha'/(16\kappa^2)$ (having taken $e^{-\phi}\simeq 1$) that can be identified with $1/(4g^2)$ where $g$ is the gauge coupling strength \cite{Polchinski:1998rq,Green:2012pqa, Uranga:2010zz}. For the photon, this is just the electric charge $e$. Using this, we have:
\begin{equation}
\frac{\alpha'}{16\kappa^2}=\frac{1}{4 e^2}
\end{equation}
This corresponds to the horizontal line in Fig. \ref{fig:boundonalpha} from where we can extract the value of electric charge present in the black hole.

\section{Discussions and conclusion}\label{resdis}
In this work, we studied the features of the KSBH in light of the recent M87* black hole observations from EHT and studies of polarization of light coming from near the black hole. The KSBH is a rotating charged black hole solution arising from string theory. As reported by the EHT, the black hole shadow at an inclination angle of $17^0$ is not completely circular with an experimental upper bound of $10\%$ on the deviation from circularity, $\Delta C$. Kerr black holes arising from general relativity cast highly circular shadows irrespective of the inclination angle leaving room for the M87* to be a non-Kerr black hole. We calculated the $\Delta C$ of the KSBH shadow using the observed M87* parameters from EHT in Sec. \ref{image}. We also found the rotation angle of the polarization of the circularly polarized light, $\Delta \Theta$, due to the axion hair of the KSBH in Sec. \ref{Pol}. Identifying the inverse string tension $\alpha'$ in terms of the electromagnetic coupling strength $e$ in the relation for $\Delta\Theta$ gives us the charge of the KSBH which can then be used to predict the $\Delta C$ of the KSBH shadow in a future precision experiment. Simultaneous observation of a constant $\Delta \Theta$ (independent of the frequency of light) and $\Delta C$ for a particular value of the spin of the black hole, its inclination angle, its mass and charge could be a signature of the KSBH and thus indirectly of the string theory. We give the expected values of $\Delta\Theta$ and $\Delta C$ for some benchmark values of the KSBH charge and its inclination angle in Table \ref{tab:prediction} for a BH spin of $a/M=0.5$. As can be seen from the table, for values of the inclination angle other than $90^0$, ($\theta\neq 90^0$), it's possible to make simultaneous observations of $\Delta\Theta$ and $\Delta C$ allowing the BH to be identified as a Kerr-Sen black hole. 

\begin{table}[h!]
\begin{center}
\begin{tabular}{|c|c|c|c|c}
\cline{1-4}
$\frac{Q}{M}$ & Inclination ($\theta$ in degrees) & Polarization Angle ($\Delta \Theta$ in degrees)  & Deviation ($\Delta C$) &\\ \cline{1-4}
 & $10$ & $1.54$ & $0.00021$ & \\
\cline{2-4}
 & 30 & 1.34 & 0.00124 & \\
\cline{2-4}
0.1 & 50 & 0.99 & 0.00282 & \\
\cline{2-4}
 & 70 & 0.53 & 0.00414 & \\
\cline{2-4}
 & 90 & 0 & 0.00463 & \\
\cline{1-4}
 & 10 & 14.79 & 0.00018 & \\
\cline{2-4}
 & 30 & 12.95 & 0.00136 & \\
\cline{2-4}
0.3 & 50 & 9.54 & 0.00310 & \\ 
\cline{2-4}
 & 70 & 5.06 & 0.00453 & \\
\cline{2-4}
 & 90 & 0 & 0.00507 & \\ 
\cline{1-4}
 & 10 & 47.56 & 0.00031 & \\
\cline{2-4}
 & 30 & 41.60 & 0.00168 & \\
\cline{2-4}
0.5 & 50 & 30.64 & 0.00380 & \\ 
\cline{2-4}
 & 70 & 16.25 & 0.00554 & \\
\cline{2-4}
 & 90 & 0 & 0.00618 & \\
\cline{1-4}
 & 10 & 119.55 & 0.00030 & \\
\cline{2-4}
 & 30 & 104.56 & 0.00242 & \\
\cline{2-4}
0.7 & 50 & 77.06 & 0.00545 & \\ 
\cline{2-4}
 & 70 & 49.81 & 0.00786 & \\
\cline{2-4}
 & 90 & 0 & 0.00874 & \\
\cline{1-4}
 & 10 & 301.29 & 0.00063 & \\
\cline{2-4}
 & 30 & 265.01 & 0.00479 & \\
\cline{2-4}
0.9 & 50 & 195.64 & 0.01059 & \\ 
\cline{2-4}
 & 70 & 103.59 & 0.01492 & \\
\cline{2-4}
 & 90 & 0 & 0.01646 & \\ 
\cline{1-4}
 \end{tabular}

\end{center}
\caption{The rotation angle of the polarization of circularly polarized light and the deviation from circularity of the BH shadows for various benchmark values of charge and inclination angle. The spin of the BH is fixed at $a=0.5$ and $\alpha'/(16\kappa^2)=1/(4e^2)$.}\label{tab:prediction}
\end{table}

%To conclude, we have calculated the deviation from circularity of the KSBH shadow and the polarization angle due to the interaction between polarized light and the axion hair of the KSBH. Simultaneous study of these two for the KSBH could be a possible means of identifying a BH as arising from the 4-dimensional heterotic string theory. \textbf{[A couple more lines here.]
There are three distinct features which can distinguish KSBH from a generic Kerr-Newman BH with axion hair:
\begin{itemize}
\item Although the charge of the KSBH is electric in nature it is sourced by the axion-photon coupling instead of the in-falling charged particles, such as electrons or positions as in the case of charged astrophysical BH like the KNBH. 

\item Owing to a different origin of charge, the KSBH can hold more of it than a KNBH. $Q/M$ for KSBH can be upto 1.4 whereas for KNBH, it cannot exceed 1.

\item In the KSBH, $\Delta \Theta$ and $\Delta C$ are correlated and the measurement of one yields the value of the other whereas in the KNBH these are independent observables.
\end{itemize}

Future, more precise, observations may be able to verify if a string solution exists in nature in the form of a Kerr-Sen black hole.

\newpage
\bibliographystyle{unsrt}
\bibliography{KS}

\end{document}